\documentclass[10pt,prb,aps,twocolumn,superscriptaddress,showpacs,amsfonts,amsmath,amssymb,showkeys,floatfix]{revtex4-1}
\usepackage{bm}
\usepackage{pstricks}
\usepackage{setspace}
\usepackage{graphicx}

\begin{document}
\title{Low-temperature spin-glass behavior in a diluted dipolar Ising system}
\date{\today}
\author{Juan J. Alonso}
\affiliation{F\'{\i}sica Aplicada I, Universidad de M\'alaga, 29071 M\'alaga, Spain}
\affiliation{Instituto Carlos I de F\'{\i}sica Te\'orica y Computacional,  Universidad de Granada, 18071 Granada, Spain}

\date{\today}

\begin{abstract}
Using Monte Carlo simulations, we study the character of the spin-glass (SG) state of a site-diluted dipolar Ising model. 
We consider systems of dipoles randomly placed on a fraction $x$ of all $L^3$ sites of a simple cubic  lattice that
point up or down along a given crystalline axis. For $x\lesssim 0.65$ these systems are known to exhibit an 
equilibrium spin-glass  phase below a temperature $T_{sg} \propto x$. At high dilution and very low temperatures, well deep in the SG phase,
we find spiky distributions of the overlap parameter $q$ that are strongly sample-dependent. We focus on spikes associated with large
excitations. From cumulative distributions of $q$ and a pair correlation function averaged over several thousands of samples
we find that, for the system sizes studied, the average width of spikes, and the fraction of samples with spikes higher than a certain threshold 
does not vary appreciably with $L$. This is compared with the behavior found for the  Sherrington-Kirkpatrick model.
\end{abstract}

\maketitle

\section{Introduction}
\label{intro}

Complex systems are present in life and social sciences, information systems, and economics. \cite{page}
In these systems, different random distributions of their microscopic constituent parts give rise to 
diverse values of some macroscopic properties.\cite{complexity} A paradigmatic model 
in statistical physics that exhibits complexity is the Sherrington-Kirkpatrick (SK) model, \cite{sk} 
where the couplings between any pair of spins is
randomly fixed to be ferromagnetic (FM) or antiferromagnetic (AF) regardless of the spin-spin distance. This model
has both quenched spatial-disorder and frustration,\cite{frustration} 
the two essential ingredients of spin glasses (SG).\cite{bookstein, bookmezard}
Its exact solution \cite{solutionSK} shows the existence 
of replica symmetry breaking (RSB):\cite{rsb} different identical replicas of a sample ${\cal J}$ with the same 
couplings may, in the thermodynamic limit, stay trapped in different states within the set of infinitely many pure states. These pure states are diverse, in the sense that they are sample-dependent. 
The distribution of the overlap $q$ between states of a given sample, $p_{\cal J}(q)$,  is 
found to be a comb-like superposition of infinitely many $\delta$-like spikes. In the macroscopic limit, after averaging 
over spatial disorder, the overlap distribution is given by $p(q)=\delta(q-q_{m}) +f(q)$ where $f(q)$ is a
non-zero smooth function for $q<q_{m}$ and zero otherwise. 

Whether the RSB picture describes correctly the behavior of  {\it realistic} spin glasses, such as  
dilute metallic alloys or concentrated insulators \cite{binderREV} is still a matter of debate.
The 3D Edwards-Anderson model (EA),\cite{ea} in which only nearest-neighbor spins interact,
is the simplest one with the essential ingredients of short-ranged SGs. No exact
solution exists for the EA model, but there is consensus on the existence of a SG phase based on
numerical simulations.\cite{oldEA}  The applicability of a RSB scenario to the SG phase of the EA model is still 
controversial.\cite{newman-stein} In the so-called \emph{droplet picture}, the SG phase is described in terms
of a unique state (paired with the one obtained by a global spin inversion) with excitations that are compact
droplets of the inverted state. \cite{droplet} According to this scenario, $p_{\cal J}(q)$ distributions do not exhibit diversity ---in the sense that  $p_{\cal J}(q)$ is independent of  $\cal J$--- and
the averaged $p(q)$ becomes a single delta function $\delta(q-q_{m})$:  $p(q)$ is said to be \emph{trivial}. 
Some  \emph{trivial-non-trivial} scenarios, between the droplet and RSB pictures, have also been proposed. \cite{kp}
Early MC simulations point to a \emph{non-trivial} scenario, \cite{kpy} but it has been found that the asymptotic 
behavior for $p(q\simeq0)$ is only reached at very large sizes even for toy droplet models. \cite{toydroplet} 

There is growing interest in the study of sample-to-sample fluctuations of $p_{\cal J}(q)$ from 
their average $p(q)$.\cite{tutti} Some recently proposed quantities give information on the height \cite{yucesoy}
and average width\cite{pcf} of spikes found in $p_{\cal J}(q)$. By MC simulations, these quantities have 
been found to be nearly size-independent for the EA model, in contradiction with RSB 
predictions. However, these results have been criticized to be far away from the asymptotic regime.\cite{critico, billoire} Some have found more useful to study the statistics of the area under $p_{\cal J}(q)$ for $q<q_{m}$.\cite{tutti,billoire}
The numerical study of SG models distinct from the SK and EA models has shed some light on the virtues and weaknesses of
these new probes for measuring diversity. \cite{middleton13, wittmann}  

Frustration in the SK and EA models comes from the competition 
between randomly distributed FM and AF couplings. However, frustration may also appear
in fully occupied systems with no quenched disorder, such as the Ising model with pure AF interactions on a FCC lattice.\cite{henley}
Dipoles packed in crystalline arrangements have frustration, exhibiting magnetic order that is strongly
dependent on the lattice structure. \cite{tisza, tisza2}  Some ferroelectrics \cite{luo} and magnetic insulators such as LiHoF$_{4}$ are known 
to be well  described by arrays of parallel Ising dipoles that behave as uniaxial ferromagnets. \cite{holmio} 
In dipolar Ising systems (DIS), dilution put together with the built in \emph{geometric frustration}  results in SG behavior. \cite{vil}  
LiHo$_x$Y$_{1-x}$F$_4$ is one example that has been extensively studied.
Experiments\cite{experimentsHO}  have found a SG phase for concentration $x=0.16$,  
and a FM phase for $x > x_{c}$ where $x_{c}\simeq 0.25$.
Recent MC simulations of systems of classical Ising dipoles placed on a fraction $x$ of the sites 
of the LiHoF$_{4}$ tetragonal lattice have found a SG phase for all $0<x\lesssim 0.25$ for temperatures below a SG transition temperature 
$T_{sg}\propto x $. \cite{tam,andresen}

Here we study a site-diluted system of $L^{3}$  dipoles, which are
placed at random on a fraction $x$ of the sites of a simple cubic (SC) lattice and point up or down along one of the principal axis.
In the limit of low concentrations details of the lattice are expected to become
irrelevant. Therefore, our model at low concentrations has direct connection  with the experimental and numerical
work mentioned above. In previous MC work \cite{ourPAD} we have calculated the entire diagram of the system and
found a SG phase for $0<x < x_{c}$ where $x_{c}\simeq 0.65$ with the SG transition temperature given by $T_{sg}(x)\simeq x$.
We found from the following evidence that the SG phase behaves marginally: (i) the mean values $q_{1}=<|q|>$ decrease
algebraically with $L$ , (ii) averaged overlap distributions of $q/q_{1}$ appear wide and independent of $L$,
(iii) $\xi_L/L$, where $\xi_L$ is a correlation length,\cite{cara} rises with $L$ at constant $T$, but extrapolates to 
finite values as $1/L\to 0$. All of this is consistent with quasi-long-range order in the SG phase,
Neither the droplet model nor a RSB scenario fit with this marginal behavior. 

The main aim of this paper is to study  whether diversity may emerge in this \emph{geometrically frustrated} model at low
temperatures and high dilution, rather deep in its marginal SG phase
by using the probes for measuring diversity, in the sense that was specified above. The paper is organized  as follows. In Sec. \ref{mm} we define the model,  give details on the parallel tempered Monte Carlo (TMC) algorithm, \cite{tempered} and 
define the quantities we compute. We present results in  Sec.~\ref{results}, 
followed by concluding remarks in Sec.~\ref{conclusion}.

\section{model, method, and measured quantities}
\label{mm}
\subsection{Models}
\label{models}
We consider site-diluted systems  of classical Ising spins  on a SC lattice.
All spins are parallel and point along the $z$ axis of the lattice. 
At each lattice site a spin is placed with probability $x$. These spins 
are coupled solely by dipolar interactions. The Hamiltonian is given by
\begin{equation}
{\cal H}=\sum_{ <i,j>}
T_{ij}\sigma_i \sigma_j ,
\end{equation}
where the summation runs over all pairs of occupied sites $i$ and $j$ except $i=j$, 
$\sigma_i=\pm 1$ on any 
occupied site $i$, 
\begin{equation}
T_{ij}=\varepsilon_a
(a/r_{ij})^3(1-3
z_{ij}^2/r_{ij}^2),
\label{T}
\end{equation} 
where $ r_{ij}$ is the distance between $i$ and $j$ sites, $z_{ij}$ is the $z$ component of $ r_{ij}$, $\varepsilon_{a}$ is an energy, and $a$ is the
SC lattice constant. In the following, temperatures shall be given in terms of $\varepsilon_a/k_B$.

Note that $T_{ij}$ values are not distributed at random, but depend only on the
orientation of vectors ${\bf r}_{ij}$ on a SC lattice. This is why DIS 
exhibit AF order at  concentrations $x > x_{c}$. \cite{tisza2, ourPAD}.
This is to be contrasted with {\it random-axes} dipolar models,\cite{rads}  in 
which Ising spins point along directions that are chosen  at random, introducing randomness on bond
strengths. 

In this paper we study DIS with  $x<<x_{c}$ for which the details of the
lattice structure are not important. Not surprisingly, in the limit of high dilution the behavior of DIS and the  
$LiHo_{x}Y_{4-x}F_{4}$ system (a well known dipolar ferromagnet for $x=1$) have been found to 
be closely related. \cite{ourPAD, andresen}

For comparison, we study also the SK model: a set of 
$N=L^3$ Ising spins $\sigma_i=\pm 1$  with  interaction energies between any pair of spins at sites $i$ and $j$ given 
by $J_{ij}\sigma_i\sigma_j$ with  $J_{ij}=\pm 1/\sqrt{N}$ chosen randomly, without bias, for all $ij$ site pairs.

\subsection{Method}

For the models described in Sec.~\ref{models} we have simulated a large number $N_{s}$  of independent {\it samples}. 
By a {\it sample}, $\cal J$, we mean a system with a given quenched distribution of empty sites for DIS 
(a quenched distribution of random couplings $J_{ij}$ for the SK model). 
The number of samples we average over is given in Tables I and II.  We have tried not to make 
$N_s$ smaller with increasing $L$. This is because statistical errors are independent  of $L$, because of non-self-averaging. 
(However, for DIS with $L=10$, we could only do $1.2 \times10^{4}$ samples. That took an Intel 8-core Xeon processor E5-2670 some $60$ years worth of CPU time).

Thermal averages come from averaging over the time range $[t_0,2 t_0]$, where $t_{0}$ is the equilibration
time.  We further average over the $N_s$  samples with different realizations of quenched disorder. 

In order to  accelerate equilibration at low temperatures in the glassy phase 
we use  a parallel tempered Monte Carlo (TMC) algorithm.\cite{tempered}
We apply the TMC algorithm as follows. We run in parallel a set of $n$ identical replicas of each sample at different 
temperatures in the interval $[T_{min},T_{max}]$ with a separation $\Delta$ between neighboring temperatures.
Each replica starts from a  completely disordered spin configuration $\{\sigma_{i}\}$.  
We apply the TMC algorithm  to any given sample in two stages. In the first stage, the $n$ replicas of the sample $\cal J$ evolve  independently for $10$ MC Metropolis sweeps.\cite{mc}  All dipolar fields throughout the system are
ufated every time a spin flip is accepted. In the second stage, 
we give any pair of replicas evolving at temperatures $T$ and $T-\Delta$ a chance to exchange states between them following standard tempering rules which  satisfy detailed balance. \cite{tempered}
These exchanges allow all replicas to diffuse back and forth from low to high temperatures 
and reduce equilibration times for the rough energy landscapes of SGs.
We find it helpful to have the highest temperature $T_{max}$ larger than 
$1.6 \times T_{sg}$. We choose $\Delta$ such that at least $30 \%$ of all 
attempted exchanges are accepted for all $T$.

For DIS we use periodic boundary conditions (PBC). 
Details of the PBC scheme we use can be found in Ref. \cite{ourPAD}.
We let a spin on an occupied site $i$ interact only with spins within an $L\times L\times L$
cube centered on $i$. In spite of the long-range nature of the dipolar
interaction, we do not perform Ewalds's summations and exclude any contributions 
from repeated copies of the lattices beyond this box. 
This introduces an error which was shown  for DIS in SC lattices to vanish as $L\rightarrow \infty$, regardless of whether 
the system is in the paramagnetic, AF or SG phase (see Appendix I in Ref.~\cite{ourPAD}). 
This result  is not applicable to an inhomogeneous FM phase that may obtain on other lattices such as in LiHoF$_{4}$.

\begin{table}[!h]

\begin{tabular}{p{0.8cm} p{1cm } p{1cm } p{1cm } p{1.4cm } p{1.4cm }}
\hline
\hline
$L$ & $T_{min}$ & $T_{max}$ & $\Delta$ & $t_{0}$ &$N_{s}$\\
\hline
$4$ &  $0.16$       & $1.60$         & $0.04$ & $10^{5}$ & $10^{5}$ \\
$6$ &  $0.16$       & $1.60$         & $0.04$ & $10^{5}$ & $1.4\times 10^{5}$ \\
$8$ &  $0.16$       & $1.60$         & $0.04$ & $2\times10^{5}$ & $10^{5}$ \\
\hline
\hline
\end{tabular}
\caption{ Simulation parameters for the SK model. The number of spins is $N=L^{3}$, $T_{min}$ 
($T_{max}$) is the lowest (highest) temperature and $\Delta$ is the temperature step in our TMC
simulations. The number of MC sweeps for equilibration is $t_{0}$. Measurements are taken 
in the time interval $[t_{0}, 2t_{0}]$.
The number of samples with different realizations of (quenched) disorder is $N_{s}$.} 
\label{tablaSK}
\end{table}

\begin{table}[!h]
\begin{tabular}{p{0.4cm} p{0.9cm } p{0.6cm } p{0.9cm } p{0.9cm } p{1cm } p{1.3cm } p{1.3cm }}
\hline
\hline
$L$ & $T_{min}$ & $T_{m}$&$T_{max}$ & $\Delta_{1}$ & $\Delta_{2}$&$t_{0}$ &$N_{s}$\\
\hline
$4$ &  $0.05$           & $0.4$ &  $1.65$      & $0.025$ & $0.05$ &$5\times10^{6}$ & $2\times 10^{5}$ \\
$6$ &  $0.05$           & $0.4$ &  $1.65$      & $0.025$ & $0.05$ &$5\times10^{6}$ & $1.1\times 10^{5}$ \\
$8$ &  $0.05$           & $0.4$ &  $1.65$      & $0.025$ & $0.05$ &$5\times10^{6}$ & $10^{5}$ \\
$10$ &  $0.075$       & $0.4$ &  $1.65$      & $0.025$ & $0.05$ &$5\times10^{7}$ & $1.2\times 10^{4}$ \\
\hline
\hline
\end{tabular}
\caption{ Same as in Table \ref{tablaSK} but for DIS with concentration $x=0.35$. In our TMC
runs for DIS we have chosen a temperature step  of $\Delta_{1}$ for the temperature interval $[T_{min},T_{m}]$,
and a bigger one, $\Delta_{2}$, for $[T_{m},T_{max}]$. }
\label{tablaDIS}
\end{table}

\subsection{ Measured quantities}
\label{meas}
Measurements were performed after two averagings: first over thermalized states
of a given sample  and second over a number $N_{s}$ of different samples. 

Given an observable $\verb|u|$, we let $u_{\cal J}=\langle~ \verb|u|~  \rangle_{T}$ stand for the thermal 
average of sample $\cal J$ and $u= [~u_{\cal J~}]_{\cal J}$ for the average over samples.

We  measure the Edwards-Anderson overlap parameter,\cite{ea}
\begin{equation} 
q=N^{-1} \sum_j \sigma^{(1)}_j\sigma^{(2)}_j ,
\label{q}
\end{equation}
where 
$\sigma^{(1)}_j$ and $\sigma^{(2)}_j$ are the spins on site $j$ of identical replicas $(1)$ and $(2)$ of a given sample. Clearly,  $q$  is a measure of the spin configuration overlap between configurations of the two replicas. 

For each sample  $\cal J$ we compute the overlap probability distribution $p_{\cal J}(q)$. The 
mean overlap distribution $p(q)$ over all replicas is defined by 
\begin{equation} 
p(q) =  [~ p_{\cal J}(q) ~]_{\cal J}.
\label{pq}
\end{equation}

We also measure the mean square 
deviations of $p_{\cal J}(q)$, from the average $p(q)$,
\begin{equation} 
\delta p(q)^{2} =  [~ \{p_{\cal J}(q) -p(q)\}^{2}~ ]_{\cal J} .
\label{dq}
\end{equation}

In order to probe for RSB behavior we focus on overlaps between states that belong to different
basins of atraction. With that aim, we compute the integrated probability functions defined by
\begin{equation} 
X_{\cal J}^{Q} =   \int_{-Q}^{Q} p_{\cal J}(q) dq,
\label{xq}
\end{equation}
\begin{equation} 
\Delta _{\cal J}^{Q} =  \big( \int_{-Q}^{Q} \{p_{\cal J}(q) -p(q)\}^{2} dq~\big)^{1/2},
\label{Deltaq}
\end{equation}
and calculate their corresponding averages $X^{Q}$ and $\Delta^{Q}$.
An advantage of working with quantities integrated over the interval
$q \in (-Q,Q)$ is that statistical errors come smaller.

Given that $X_{\cal J}^{Q}$ is a ($\cal J$-dependent) random variable,
it makes sense to explore how this variable is distributed. Following Reference [\onlinecite{billoire}], we 
define its cumulative distribution $\Pi_{c}^{X} (z)$ as  
the fraction of samples having $X_{\cal J}^{Q} < z$.

Yucesoy  {\it et al.} \cite{yucesoy} have proposed very recently an observable
that is sensitive to spikes in the overlap distributions $p_{\cal J}(q)$ of individual samples. 
They consider the maximum value of $p_{\cal J}(q)$ for $q \in (-Q,Q)$,
\begin{equation} 
\\\widetilde{p}_{\cal J}^{~Q} =  \max\{~p_{\cal J}^{Q}(q) : |q|<Q~\},
\label{peaksq}
\end{equation}
and count a sample as \emph{peaked} if $\widetilde{p}_{\cal J}^{~Q}$ exceed some specified value.
We  compute the cumulative distribution $\Pi_{c}^{\widetilde p} (z)$ of $\widetilde{p}_{\cal J}^{~Q}$
as  the fraction of samples having $\widetilde{p}_{\cal J}^{~Q} < z$.

In previous papers \cite{pcf} we have obtained additional information on the shape and width 
of spikes  from a \emph{pair correlation function}. 
Let $f_{\cal J}(q_1,q_2)\equiv   p_{\cal J}(q_1)p_{\cal J}(q_2) $, 
\begin{equation}
G_{\cal J}^{~Q}(q)  = \int_{0}^{Q} \int_{0}^{Q} dq_1 dq_2\; \delta (q_2-q_1-q)f_{\cal J}(q_1,q_2),
\label{GG}
\end{equation}
and let $G^{Q}(q)$ be the average of $G_{\cal J}^{~Q}(q) $ over samples. 
We compute the normalized function
\begin{equation}
g^{Q}(q)  = G^{~Q}(q)/  \int_{-Q}^{Q} G^{Q}(q) ~dq,
\label{g}
\end{equation}
which is the conditional probability density that  $q=q_2-q_1$, given that $q_1,q_2\in (0,Q)$. 
Note that $g^{Q}(q)$ is largest at $q=0$, and that $g^{Q}(q)=g^{Q}(-q)$, since  $p_{\cal J}(q)=p_{\cal J}(-q)$.
It makes sense to define the width of $g^{Q}(q)$ as 
\begin{equation}
w^{Q} =1/g^{Q}(0),
\label{defw}
\end{equation}
which is a measure of pattern thermal fluctuations for $|q|<Q$.

An additional interpretation of $g^{Q}(q)$ is possible for sufficiently small $T$ ($T\lesssim 0.4T_{sg}$, roughly)
so that individual spikes are clearly discernible.
Assume, in addition, that $Q$ is sufficiently small so that contributions from samples with more than one 
spike in the $0<q<Q$ range is negligible. Then, (i) finding on each sample one such spike, if there is one, 
(ii) calculating the self-overlap of such spike with a copy of itself shifted by a distance $q$ 
(iii) adding the resulting function of $q$ over all samples, 
and (iv) normalizing, gives $g^{Q}(q)$. To that extent, $g^{Q}(q)$ stands for an average over all spikes on the $0<q<Q$ range. 
In  Ref.~\cite{pcf} we have also shown that 
if the width of spikes does not vary over different samples, then
$g^{Q}(q)$ is, for large systems, twice as wide as spikes are.

\subsection{Equilibration times}

\begin{figure}[!t]
\includegraphics*[width=80mm]{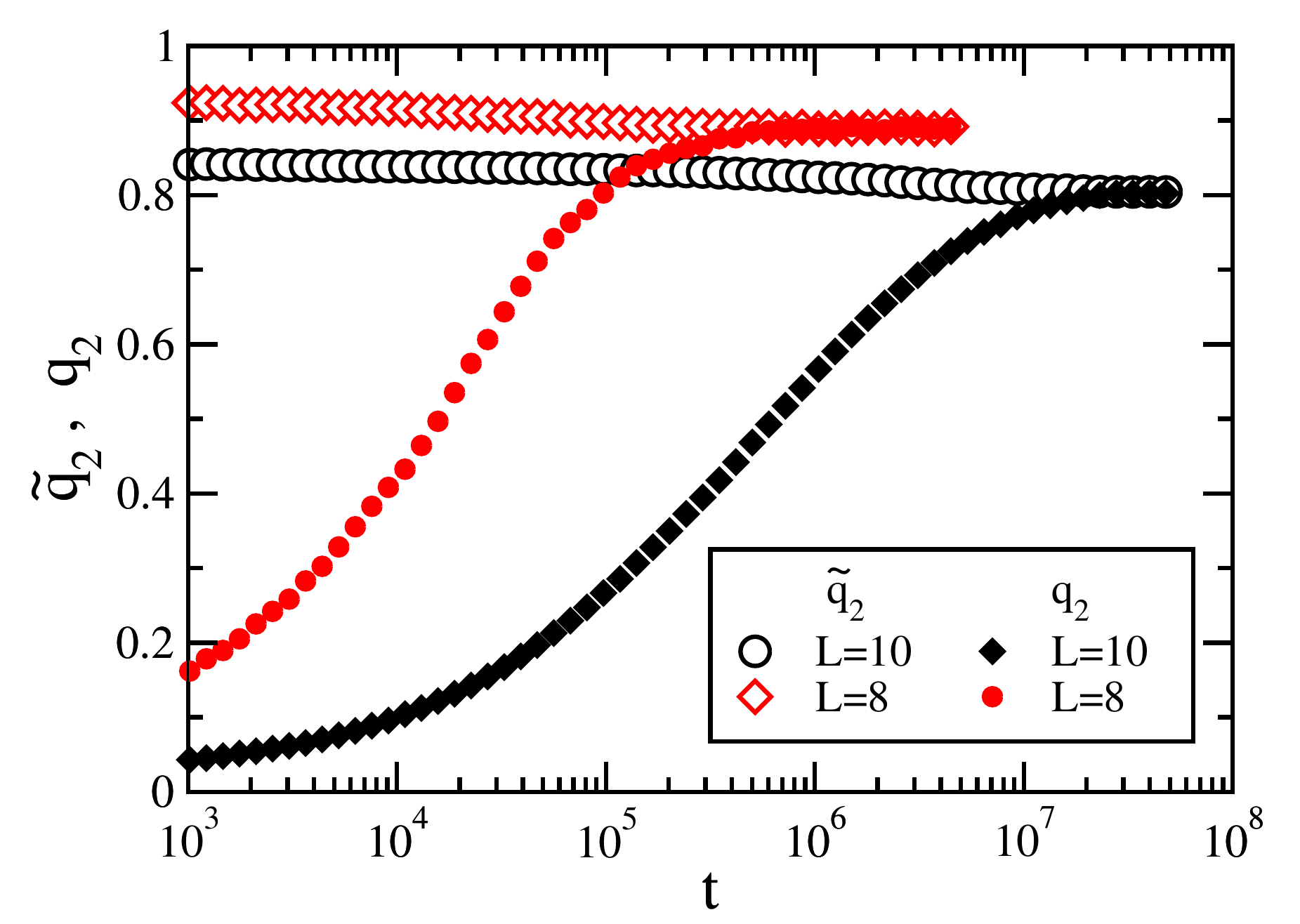}
\caption{(Color online)
Semilog plots of $\widetilde q_2(t_0,t)$ and $q_2$ vs time $t$ (in MC sweeps) for DIS systems with concentration $x=0.35$ running
at the lowest temperature $T_{n}$ for the values of $L$ indicated in the figure. $T=0.075$ ($T=0.05$) for $L=10$ ($L=8$).  $q_2$ is obtained from averages of $ q^2$ over time,  starting at $t=0$ from an initial random spin configuration. Here, $t_0=5 \times10^7$ MC sweeps for $L=10$, and $5 \times10^6$ MCS for $L=8$. Data points at time $t$ stand for an average over a time interval $[t,1.2t]$, and
over $10^{3}$ samples.
}
\label{tiempos}
\end{figure}

We now explain how we make sure that  thermal equilibrium is reached before we start taking
measurements. To this end, some quantities are next defined.
First, a pair of identical replicas of a given sample are allowed to evolve independently in time, 
starting at $t=0$ from two uncorrelated random spin configurations. Let $q_{t}$ be the overlap between the
configurations of the two identical replicas at time $t$. In addition, let $q_{2}(t)$  be the average 
of $q_{t}^{2}$ 
over all samples.  During equilibration $q_{2}(t)$
is expected to increase up to its equilibrium value.
Semilog plots of  $q_{2}(t)$ versus $t$ displayed in Fig.~\ref{tiempos} for $x=0.35$, $L=8, 10$ and the lowest used temperatures show
that a stationary value $q_{2}$ is reached only after some millions of MC sweeps. 

In order to check whether this stationary value $q_{2}$ is an equilibrium one, we define a second overlap,
$\widetilde q_{t}$, 
not between configurations of pairs of identical replicas at the same time $t$, but
between spin configurations of a single replica taken at two different times $t_{0}$ and $t_{1}=t_0+t$ of the same MC run,
\begin{equation} 
\tilde q_{t}(t_0)=N^{-1} \sum_j \sigma_j(t_0)\sigma_j( t_0+t).
\label{phi0}
\end{equation}

Let $\tilde{q}_2(t_0,t)$  be the average of $(\widetilde q_{t}(t_0))^{2}$ 
over all samples. 
Suppose thermal equilibrium is reached long before time $t_0$ has elapsed. 
Then, $\tilde{q}_2(t_0,t)$ and $q_2(t)$ should tend towards a common value $q_{2}$ as  $t \rightarrow  t_0$. 
Plots of $\tilde{q}_2(t_0,t)$ vs $t$ are shown in Fig.~\ref{tiempos} for $t_0=5\times 10^7$ MC sweeps ($5\times 10^6$ MCS sweeps) for $L=10$ ($L=8$)  for the same values of $x$ and $T$ as for $q_{2}(t)$. Note  that both quantities, $\tilde{q}_2(t_0,t)$ and $q_2(t)$, do become approximately equal when $t \rightarrow  t_0$. In order to obtain equilibrium results, we have always chosen sufficiently large values  of $t_0$ to make sure that  $\tilde{q}_2(t_0,t)\approx q_2(t)$ for  $t\gtrsim t_{0}$. 
In our simulations, we let each system equilibrate for a time $t_{0}$ and 
take averages over the time interval $[t_{0},2t_{0}]$. All values of $t_0$ and $N_s$ are given in Table II. 

\begin{figure}[!t]
\includegraphics*[width=80mm]{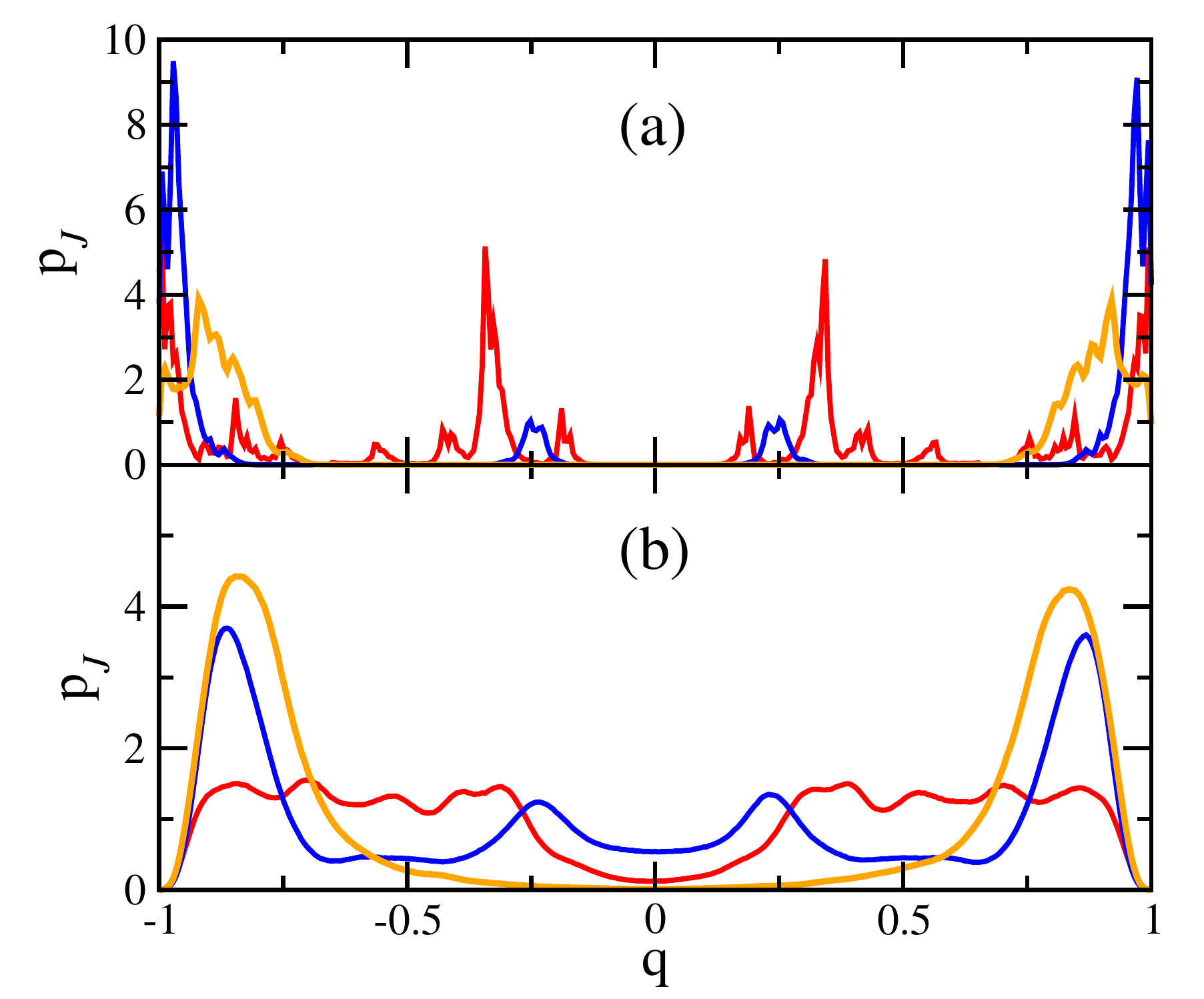}
\caption{(Color online) 
(a) Overlap distributions $p_{\cal J}(q)$ for DIS systems with $L=10$,  $x=0.35$ and $T=0.1$ for three samples with different realizations
of disorder. For each sample we collect values of $q$ 
over $5 \times 10^{7}$ MC sweeps. (b) The same as in (a) but for $T=0.25$ for the same sample set. Recall that the transition
temperature is $T_{sg}=0.35$.}
\label{pejotas}
\end{figure}

It has recently been shown that equilibration times increase with the  roughness of the free-energy landscape of each individual sample. \cite{yuce2} Numerous spikes in overlap distributions $p_{\cal J}^{~Q}$ are the signature of samples that have numerous minima in their free-energy landscape.
Visual inspection of overlap distributions of  samples like the ones shown in Fig.~\ref{pejotas} shows fairly symmetric $p_{\cal J}^{~Q}$ curves even though some of them have several spikes. Then, our stringent equilibration criterion suggests that nearly all the samples are well equilibrated.

\section{RESULTS} 
\label{results}

\subsection{Overlap distributions}
\label{od}

As it has been found for other SG models, it is interesting to examine individual samples of DIS.  In Fig.~\ref{pejotas}(a) we plot  $p_{\cal J}(q)$ versus $q$ for 
different samples at temperature $T/T_{sg}\simeq 0.3$. At
 this low temperature, some $p_{\cal J}(q)$ display well defined spikes centered 
on small $q$ values, which seems to vary randomly from sample to  sample. Qualitatively similar distributions have been observed for the EA and SK models. \cite{yucesoy, pcf, aspelmeier} It is clear that these inner peaks (that is, peaks away from $q \approx \pm 1$) come from overlaps between states that belong to different basins of attraction. The main aim of this paper is to extract statistical information for these \emph{cross-overlap} (CO) spikes situated on the interval  $q \in (-Q,Q)$. 
Similar plots for higher temperatures (see Fig. \ref{pejotas}(b) for $T/T_{sg}\simeq 0.7$), show that thermal fluctuations render individual spikes not clearly discernible. Then, in order to explore 
well within the SG phase, we have chosen the lowest temperature in our TMC simulations to be  $0.2~T_{sg}$. We report most of our results for $T\le0.4~T_{sg}$. We have also chosen a concentration  $x=0.35$ which is far below the threshold for the AF phase. Both low temperatures and low concentrations result in large equilibrium times $t_{0}$. In addition, in order to obtain good sample statistics we need to simulate a large number $N_{s}$ of samples for all system sizes studied. All of this has restricted us to deal with relatively modest system sizes in our simulations. The simulation parameters are  given in Tables I and II.

\begin{figure}[!t]
\includegraphics*[width=75mm]{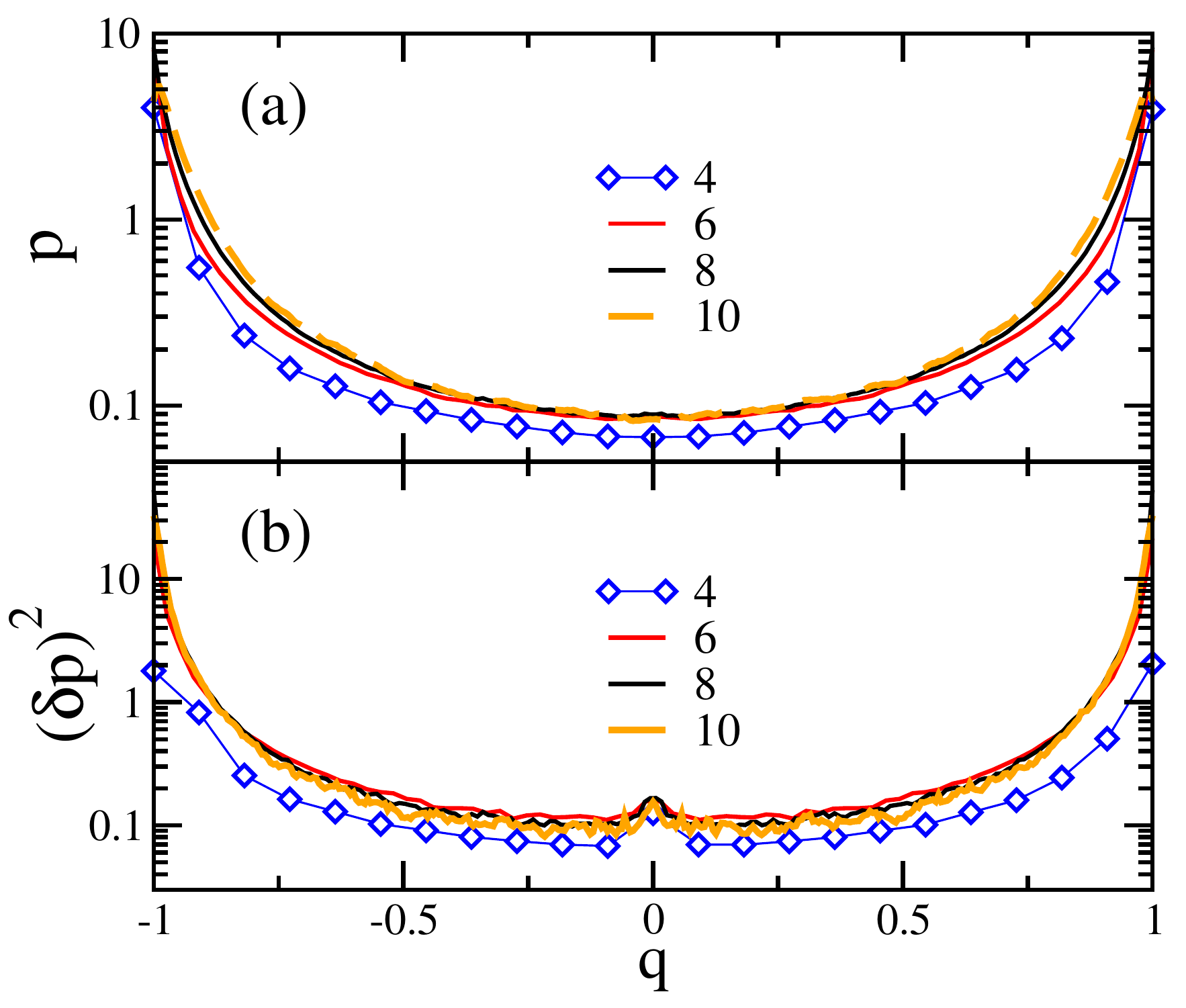}
\caption{(Color online) (a)
Plots of the averaged distribution $p(q)$ versus $q$ for DIS systems with 
$x=0.35$, $T=0.1$, and the values of $L$ shown. (b) Same as in (a), but 
for $(\delta p)^{2}$, the mean square deviations of $p_{\cal J}(q)^{2}$ away from $p(q)$ over all $\cal J$ samples.}
\label{p_delta}
\end{figure}

Figure \ref{p_delta}(a) shows the sample-averaged overlap distribution $p(q)$ for DIS
at $T=0.1$. At this low temperature, $p(q)$ exhibits two large peaks at $\pm q_{m}$ with $q_{m}\approx1$ and
a relatively flat plateau with $p(0)\ne0$ in the region $q \in (-Q,Q)$ for, say, $Q\approx 1/2$. 
The non-zero  $p(0)$ value   does not change with $L$ for the system sizes studied. 
This behavior, known for the SK and EA models,\cite{solutionSK,kpy} is in contradiction with the droplet picture of SGs, for 
which $p(0)$ vanishes as $L^{-\theta}$.\cite{droplet}

\begin{figure}[!b]
\includegraphics*[width=85mm]{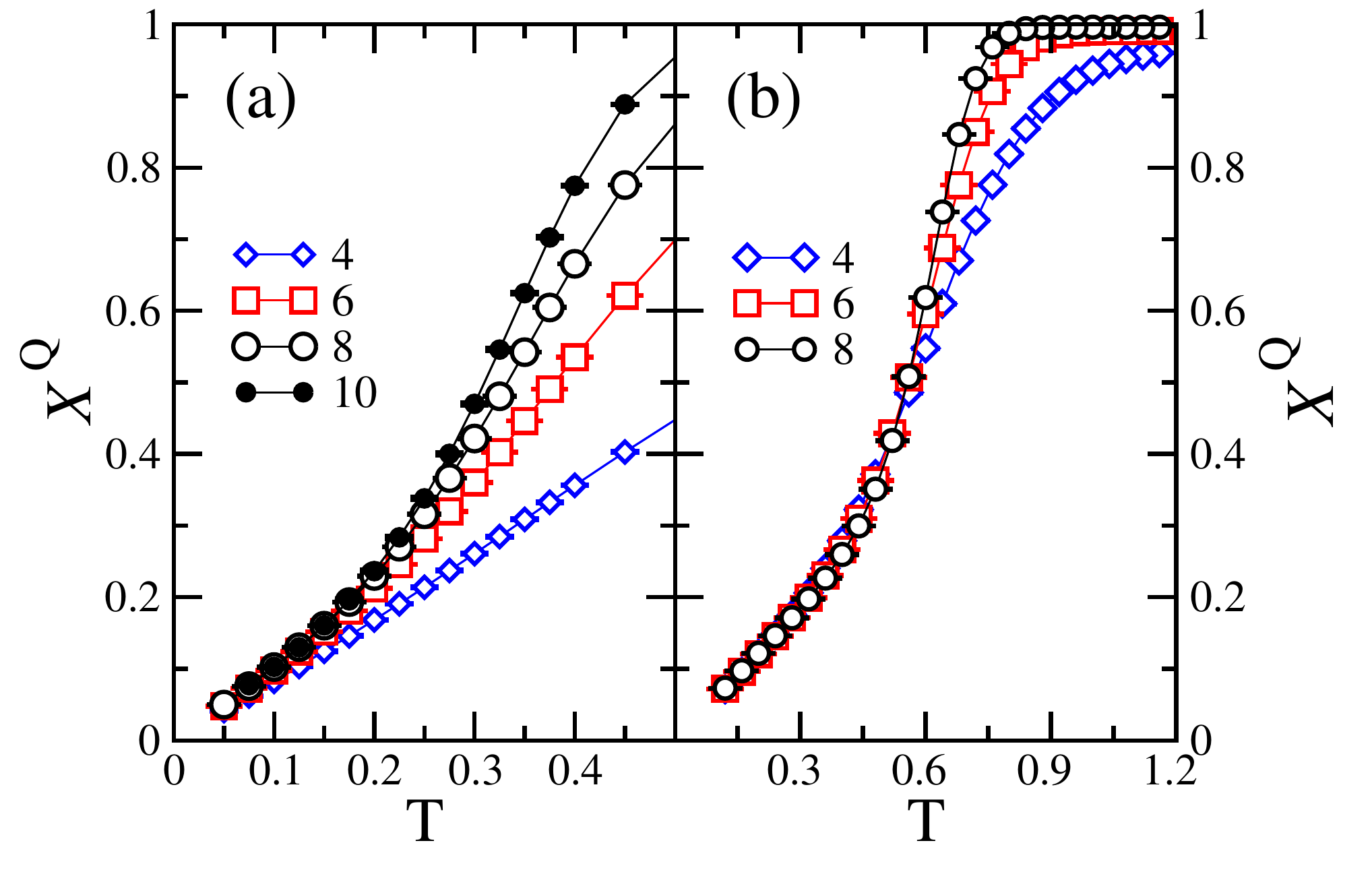}
\caption{(Color online) (a) Plot of $\it{X}^{Q}$ versus T for DIS systems with $x=0.35$, $Q=1/2$ and the values of 
$L$ indicated in the figure. (b) The same plots for the SK model, with $Q=1/2$ and the values of $L$ shown. 
In both panels, error bars are smaller than symbol sizes.}
\label{pvsT}
\end{figure}

Plots of  $(\delta p)^2$ vs $p$ are shown for DIS at the same temperature 
in Fig. \ref{p_delta}(b).
$\sqrt{(\delta p)^2}$, a measure of deviations of $p_{\cal J}(q)$  from the average $p(q)$, is clearly
greater than $p$ for $q \in (-1/2,1/2)$ indicating lack of self-averaging. More interestingly, $(\delta p)^2$ 
does not vary appreciably with $L$. This is at odds with the behavior found in MC simulations for the 
SK model  for which $(\delta p)^2 \propto L$ for $T\le0.5~T_{sg}$. \cite{ourPAD}
Recall that in the RSB scenario, one expects that $p_{\cal J}(q)$ exhibits with many sharp spikes in the region 
$q \in (-Q,Q)$ that become $\delta$-like functions as
$L$ increases, resulting in a diverging $(\delta p)^2$ for macroscopic systems.

\subsection{Integrated overlap distributions}
\label{iod}
Here we consider  averages of both $p$ and $(\delta p)^2$ over $q \in (-Q,Q)$. This allows us to focus on the 
contributions of CO spikes and, in addition, to reduce 
 statistical noise if $Q$ is not too small.  Plots of   $X^{Q}$, the sample-averaged area under CO spikes, versus $T$
 are shown for $Q=1/2$  in Figs. \ref{pvsT}(a)  and \ref{pvsT}(b) for DIS and SK models respectively. In both cases, $X^{Q}$ is, as far as we can see,  size 
 independent at temperatures well below $T_{sg}$.  We obtain quantitatively similar results (not shown) for $Q=1/4$.  
 This is a strong piece of evidence against
 the validity of the droplet picture, for which $X^{Q}$ is expected to vanish as $TL^{-\theta}$.\cite{droplet} A similar behavior 
 has been found for the EA model in several MC simulations.\cite{kpy,pcf} However, it has been argued that strong 
 finite-size effects may mask the asymptotic behavior at the system sizes
 currently available to MC simulation.\cite{toydroplet} Finally, we note that in Figs.~\ref{pvsT}(a) and (b) $X^{Q}$ seems to vanish
 as $T\to0$  (as was long ago predicted for the SK model). \cite{vanni}
 
\begin{figure}[!b]
\includegraphics*[width=85mm]{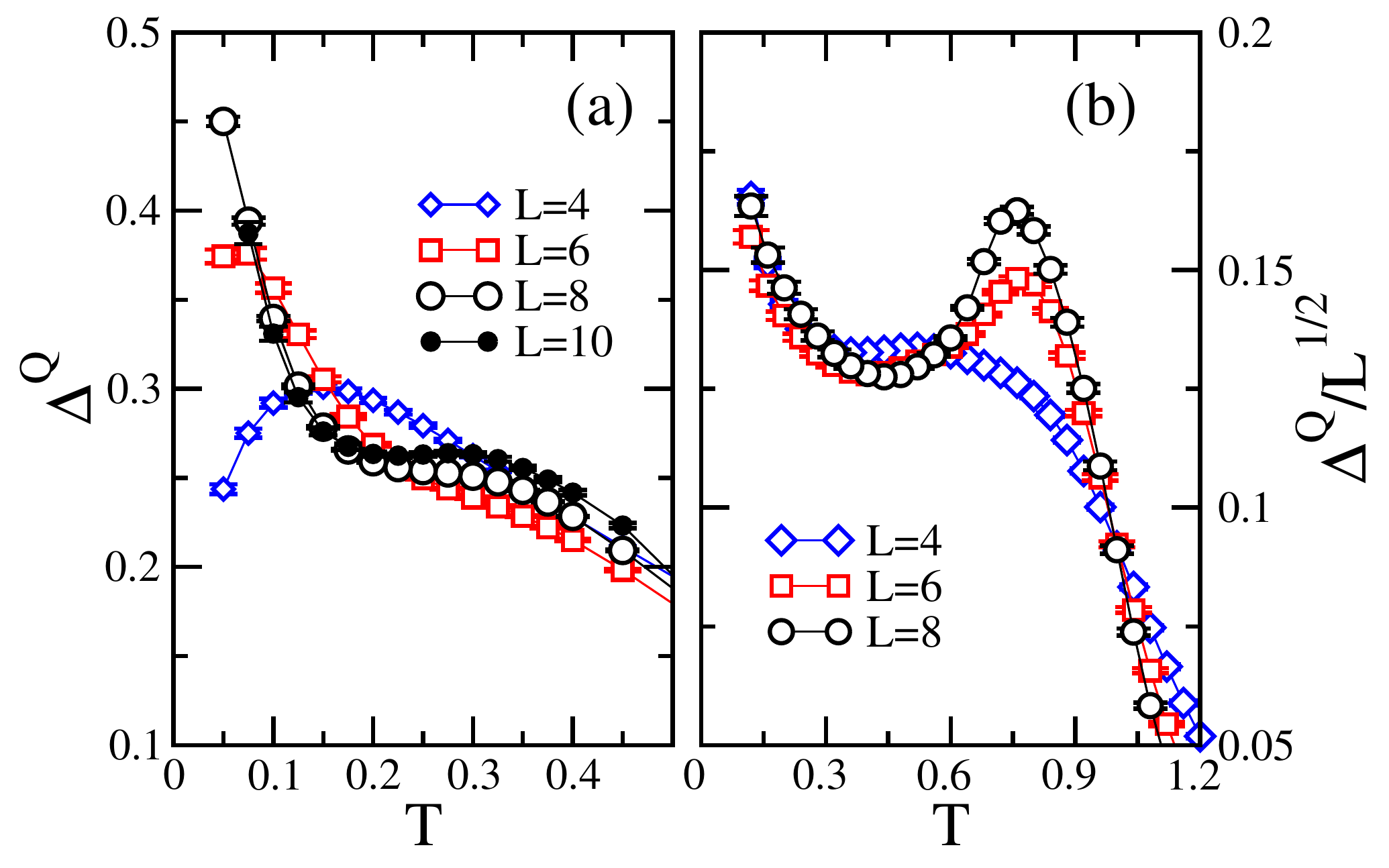}
\caption{(Color online) (a) Plot of $\Delta_{Q}$ versus T for DIS systems with $x=0.35$, $Q=1/2$ and the values of 
$L$ indicated in the figure. (b) Plot of  $\Delta_{Q}/L^{1/2}$ versus $T$ for the SK model, with $Q=1/2$. }
\label{DeltavsT}
\end{figure}

Plots of $\Delta^{Q}$ vs $T$ for DIS  in Fig.~\ref{DeltavsT}(a) show the presence of 
finite size effects.  Note that for small sizes, $\Delta^{Q}$ increases as $T$ decreases only up to
$T=0.15$ ($T=0.075$) for $L=4$ ($L=6$). We return to this point in Section \ref{pcf}.  
More interestingly, curves for larger sizes ($L\ge8$) give
a strong indication that   $\Delta^{Q}$ does not diverge as $L$ increases. This result is in contradiction with a RSB scenario and 
is in sharp contrast with the behavior  exhibited in Fig.~\ref{DeltavsT}(b) for the SK model, 
for which $\Delta^{Q}$ increases with $\sqrt{L}$  at low temperatures. It is worth mentioning that $\Delta^{Q}$
differs qualitatively from  the average $\it{X}_{2} \equiv[ ~(X_{\cal J}^{Q})^{2}~]_{\cal J}$ when $Q$ is not very small. \cite{pcf} 
$ \it{X}_{2}$ has been investigated in detail in several papers and it is known to be size independent for both the EA and SK models.
\cite{tutti}

\begin{figure}[!t]
\includegraphics*[width=65mm]{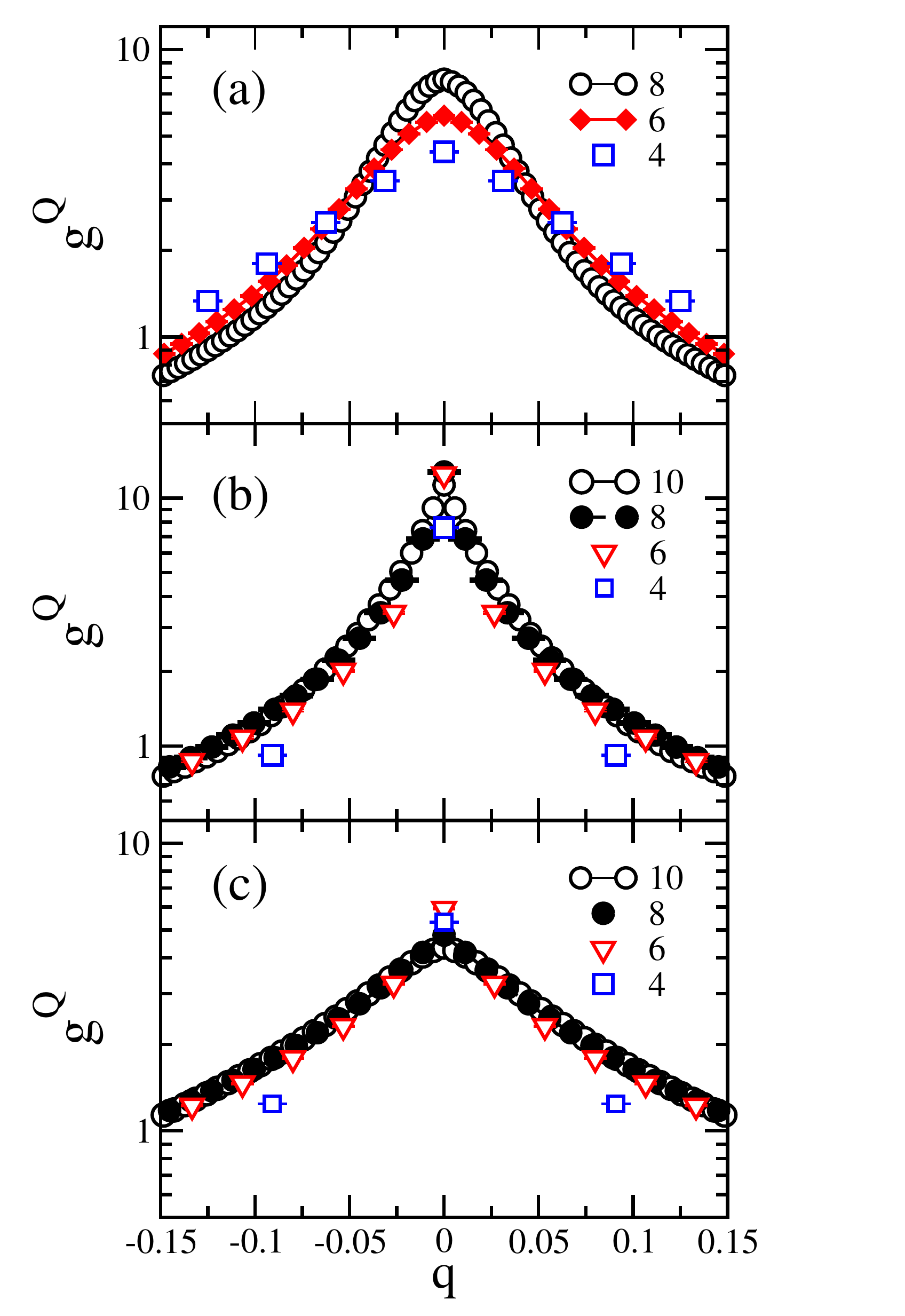}
\caption{ (Color online):
(a) Plots of $g^Q$ vs $q$ for the SK model at $T=0.2$ with $Q=1/2$ and the values of $L$ shown. (b) Same plot as for (a) but for DIS systems with $x=0.35$ at $T=0.075$. (c) Same plot as for (b) but 
for $T=0.125$.}
\label{gqtrio}
\end{figure}

\subsection{Pair correlation functions}
\label{pcf}
Data points for the pair correlation function $g^{Q}$ for $Q=1/2$ are shown in  
Fig.\ref{gqtrio}(a) and (b) for the SK model at $T=0.2$  and DIS at $T=0.075$ respectively.
Note that these temperatures are such as $T/T_{sg}\simeq0.2$ in both cases. We find curves that are rather 
pointed with widths clearly smaller than $Q$. 
We obtain similar results for $Q=1/4$. Data for DIS in Fig.~\ref{gqtrio}(b) do not exhibit any significant size dependence.
In contrast, $g^{Q}$ curves for SK in  Fig. \ref{gqtrio}(a) become sharper as $L$ increases. This result for the SK model 
is  as expected for a RSB scenario. In the RSB solution, $p_{\cal J}(q)$  is made of several cross-overlap spikes 
that for small values of $q$ become $\delta$-functions in the macroscopic limit, and densely fill the interval $q\in(-Q,Q)$. In striking contrast,
our result for DIS suggest that the number of SG states do not grow with $L$ at finite low temperatures.

Some people have argued that comparing data for different models (EA and SK) at the same value of $T/T_{sg}$ is not
meaningful. They find more appropriate to make such a comparison at temperatures for which $X^{Q}$ values are the same.\cite{critico} 
We follow this recipe and compare the data shown in Fig.~\ref{gqtrio}(c) for DIS at $T=0.125$ and in Fig.~\ref{gqtrio}(a) for the SK model at $T=0.2$. 
Apart from the fact that $g^{Q}$ becomes narrower as $T$ decreases, we do not notice any qualitative difference.

\begin{figure}[!t]
\includegraphics*[width=87mm]{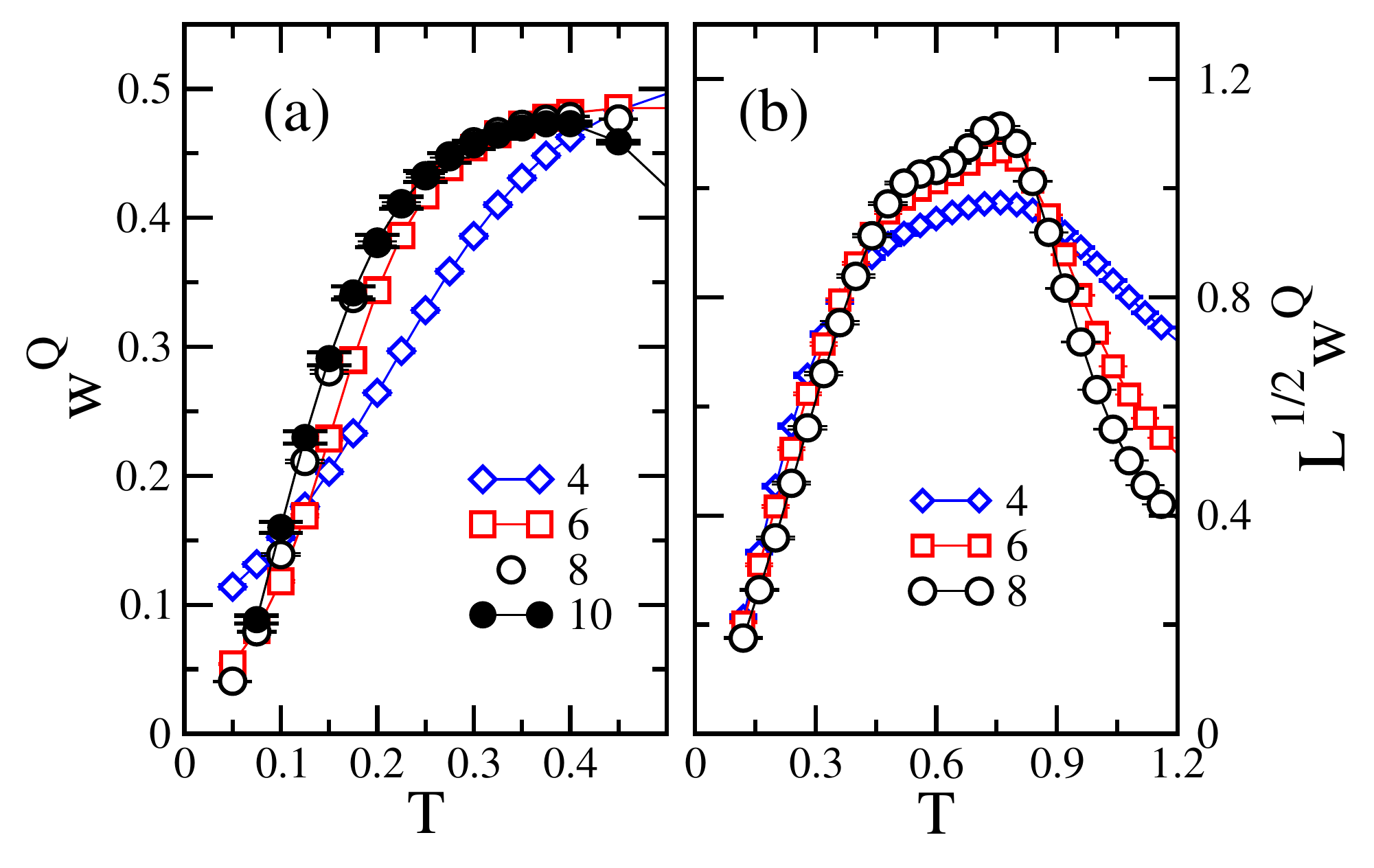}
\caption{ (Color online)
(a) Plots of $w{Q}$ vs $q$ for DIS systems with $x=0.35$  for $Q=1/2$ and the values of $L$ shown. (b) Plots of $L^{1/2}w{Q}$ vs $q$ for (a) but for the SK model, $Q=1/2$ and the values of $L$ shown. In both panels all error bars are smaller than symbol sizes.}
\label{widths}
\end{figure}

It is interesting to note that spikes, even though they have non-zero widths $w$ at finite $T$, may not be discerned
in $p_{\cal J}(q)$ distributions of very small systems. Note that the minimum appreciable value of $q$ for systems of $N$ spins is
given by $\Delta q=2/N$. Thus, finite size effects are expected to come at very low $T$ when $w \lesssim \Delta q$.  This seems
to be the case for the data shown in Figs.~\ref{DeltavsT}(a) and \ref{widths}(a) for  $L=4$ ($L=6$) and $T \lesssim 0.15$ ($T \lesssim0.075$).

Plots of $w^{Q}$ versus $T$ are shown in   Figs.~\ref{widths}(a) and ~\ref{widths}(b) for DIS and the SK model respectively. 
$w^{Q}$ appear in Fig.~\ref{widths}(a) to be size independent at least for $L\ge8$. 
This points to  finite widths for CO spikes  in the $L\rightarrow\infty$ limit for low (but finite) temperature.
On the other hand $w^{Q}$ values for the SK model displayed in Fig.~\ref{widths}(b) appear to vanish as $1/L^{1/2}$ as $L$ increases 
at least for $T\lesssim 0.4$, in agreement with the RSB picture.

\begin{figure}[!b]
\includegraphics*[width=85mm]{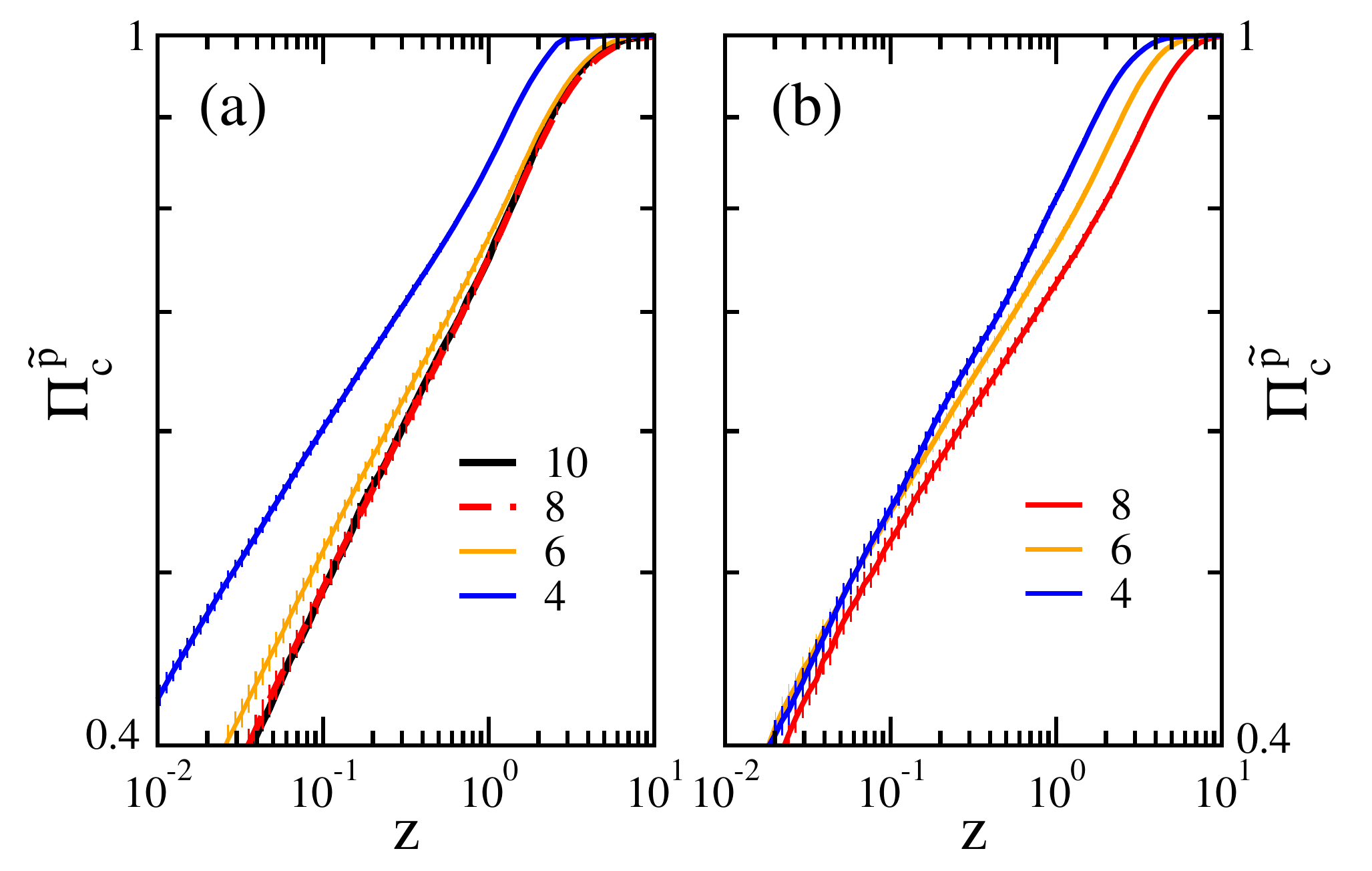}
\caption{ 
 (Color online) (a) Plot of the cumulative distribution $\Pi\{ \widetilde{p}^{~Q}_{\cal J}\}$ versus $x$ for DIS systems with $x=0.35$, for  $Q=1/2$, $T=0.1$ and the values of  $L$ indicated in the figure, where $\widetilde{p}^{Q}_{\cal J}$ is the maximum value of  $p_{\cal J}(q)$ over the interval $-Q\le q\le Q$.  (b) The same plot as in (a) but for the SK model, for  $Q=1/2$, $T=0.16$ and the values of $L$ shown.} 
\label{pijotas}
\end{figure}

\subsection{Cumulative distributions}
\label{cumul}

\begin{figure}[!t]
\includegraphics*[width=85mm]{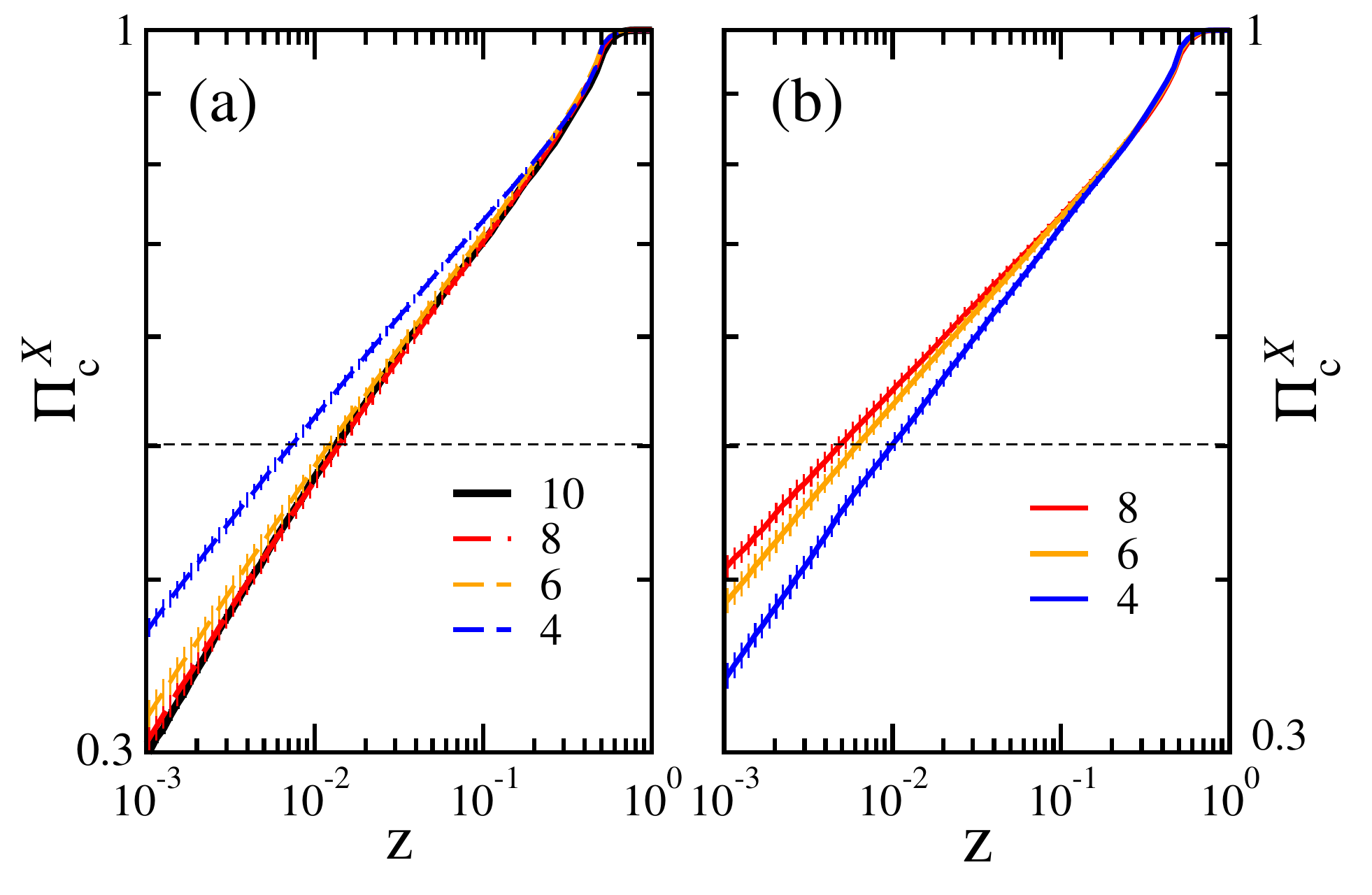}
\caption{ 
 (Color online) (a) Plot of the cumulative distribution $\Pi\{X^{Q}_{\cal J}\}$ versus x for DIS systems with $x=0.35$, for  $T=0.1$ and the values of 
$L$ indicated in the figure. (b) The same plots as in (a) but for the SK model, for $T=0.16$.}
\label{piequis}
\end{figure}

As interesting as they could be, pair correlation functions (PCFs) do only give information on how spiky 
sample distributions $p_{\cal J}(q)$ are in the  $(-Q,Q)$ region. However, PCFs do not give any 
information about the height of spikes located there. Following seminal work by Yucesoy {\it et al.} \cite{yucesoy} we study here
$\Pi_{c}^{\widetilde p} (z)$, the fraction of samples without any spike in $(-Q,Q)$ with height  
larger than $z$. Plots of $\Pi_{c}^{\widetilde p}$ versus $z$ for DIS at $T=0.1$ are shown in Fig.~\ref{pijotas}(a). They give a
strong indication that  $\Pi_{c}^{\widetilde p}$ reach a size-independent shape for  $L\ge8$ for a wide range of values of $z$.
In order to check for the robustness of our $\Pi_{c}^{\widetilde p}(z)$ values, we have grouped all available samples in $K$ 
ensembles of $10^3$ samples  each, calculated $\Pi_{c}^{\widetilde p}$ for each ensemble $k=1,...,K$, and obtained the standard deviation (SD) of
the $K$ resulting values. Tiny vertical bars in Figs.~\ref{pijotas} and \ref{piequis} stand for such SDs. Plots in Fig.~\ref{pijotas}(a) are to be compared with the ones
displayed in Fig.~\ref{pijotas}(b) for the SK model. Note in Fig.~\ref{pijotas}(b) that, at least for $z\gtrsim 0.5$, $\Pi_{c}^{\widetilde p}(z)$ clearly decreases as $L$
increases, indicating a proliferation of high spikes for larger sizes. This is as expected for the RSB picture, for which CO spikes become $\delta$-like functions in the thermodynamic limit. 

Finally, we report results on how the random variable $\it{X}_{\cal J}^{Q}$ is distributed. 
Previous MC simulations have found a similar behavior for the EA and SK models when
dealing with the cumulative distribution $\Pi_{c}^{\it X} (z)$ of quantity $z=\it{X}_{\cal J}^{Q}$. \cite{yucesoy, billoire}
The mean-field theory of the SK model offers precise predictions on $\Pi_{c}^{\it X} (z)$ and their moments. \cite{tutti}
In particular, $\Pi_{c}^{\it X} (z)$ is found to follow a power law for small $z$. For small values of $Q$,\cite{mezard} 
$\Pi_{c}^{\it X} (z) \propto z^{\it y}$, where $\it y$ stands here for $\it{X}^{Q}$. Some people have found it useful to study the \emph{median} of the cumulative distribution, which is predicted to reach a non-zero value in the thermodynamic limit in the RSB picture but vanishes for the droplet model.  Log-log plots of  $\Pi_{c}^{\it X}$ versus $X$ for DIS displayed in Fig.~\ref{piequis}(a) show curves with power-law behavior for small $z$. Data do not show any significant deviation for sizes $L\ge8$. The \emph{median} (marked by the crossings points of the curves with the horizontal dotted line in the figure)
decreases as $L$ increases reaching a non-zero value. The counterpart plots for the SK model are shown in Fig.~\ref{piequis}(b). In agreement with previous MC work on the SK model, \cite{billoire,wittmann} we find strong
finite-size effects. However, curves seem to converge to some limiting curve as $L$ increases.\cite{billoire} Note that, in 
contrast with DIS, the \emph{median} of $\Pi_{c}^{\it X}$ increases as $L$ increases. 
All our results for $\Pi_{c}^{\it X}$ for both the SK model  and DIS are not in contradiction with a RSB scenario.

\section{Conclusions and discussion}
\label{conclusion}

By tempered Monte Carlo calculations, we have studied the low temperature behavior of a diluted system of classical dipoles placed on a SC lattice. These dipoles are Ising spins randomly placed on a fraction
$x$ of all lattice sites and point up or down along a common crystalline axis. 

Previous MC studies \cite{pcf} for this model have
provided  strong evidence for the existence of a SG phase for $x\lesssim 0.65$ with a SG transition temperature 
$T_{sg}(x) \simeq x$. The SG phase was then found to have quasi-long range 
order, as in the 2D-XY model. \cite{xy} Neither the droplet model nor a RSB scenario fit with
this marginal behavior.  Despite the existence of this \emph{soft} SG order, we find in our simulations
spiky overlap distributions $p_{\cal J}(q)$ that are strongly sample-dependent, as previously
found in simulations for the EA and the SK models.\cite{yucesoy,pcf,aspelmeier}

We have studied  the statistics of $p_{\cal J}(q)$ for  $q\in(-Q,Q)$ using some 
recently proposed observables. \cite{yucesoy,pcf,billoire}
We find that $p(q)$ and $\delta p(q)$ (as well as their integrated counterparts $X^{Q}$ and $\Delta ^{Q}$)
do not vary appreciably with $L$.

From a suitable defined pair correlation function \cite{pcf} we compute an averaged width $w^{Q}$ 
that appears to remain finite as $L$ increases. Complementary to this result, we find that the 
fraction of samples with spikes higher than a certain threshold does not vary appreciably with $L$. 
All of this points to finite width for CO spikes in the $L\to\infty$ limit at  low temperatures. 

Our results are in clear contradiction with droplet model predictions. On the other hand, a direct comparison of our
data with MC data obtained for the SK model shows that crucial RSB predictions are also at odds 
with some of our results. It is noteworthy that the findings enumerated above for DIS are strikingly similar to the 
ones found in previous MC work for the EA model, \cite{yucesoy,pcf,billoire,wittmann} though finite  size
effects have been reported to be strong for the EA model for the available
systems sizes in MC simulations.

\section*{Acknowledgements}
We are indebted to the Centro de Supercomputaci\'on y Bioinform\'atica and to the 
Applied Mathematics Department both at University of M\'alaga, and to 
Institute Carlos I at University of Granada  for much computer time in clusters Picasso, Atlantico, and Proteus.
Funding from the Ministerio de Econom\'ia y Competitividad of Spain, through Grant No. FIS2013-43201-P,  is gratefully acknowledged.
We are grateful to J. F. Fernandez for helpful comments. 
We are also indebted to  B. Alles and R. Roa  for kindly reading the manuscript.

\end{document}